\documentstyle[12pt,aasms4]{article}
\tightenlines
\begin{document}
\title{The Angular Two-Point Correlation Function for the FIRST Radio Survey}

\author{Catherine M. Cress, David J. Helfand}
\affil{Department of Astronomy, Columbia University, 538 West 120th Street, 
New York, NY 10027}
\author{Robert H. Becker}
\affil{Department of Physics, University of California, Davis, CA 95616 and}
\affil{Institute of Geophysics and Planetary Physics, Lawrence Livermore 
National Laboratory, Livermore, CA 94450}
\author{Michael D. Gregg}
\affil{Institute of Geophysics and Planetary Physics, Lawrence Livermore 
National Laboratory, Livermore, CA 94450}
\author{Richard L. White}
\affil{Space Telescope Science Institute, 3700 San Martin Drive, Baltimore, MD 21218}
----------------------------------------------------------------------------
\begin{abstract}
The FIRST (Faint Images of the Radio Sky at Twenty Centimeters) survey
now covers 1550 $\rm deg^{2}$ of sky where $07^{h}16$ $^{<}_{\sim}$ $\alpha$
$^{<}_{\sim}$ $17^{h}40$ and 
$28^{\circ}.3$ $^{<}_{\sim}$ $\delta$ $^{<}_{\sim}$ $42^{\circ}$. This 
yields a catalog of 138,665 sources above the survey threshold of 1 mJy,
about one third of which are in double-lobed and multi-component sources. 
We have used these data to obtain the first high-significance measurement
of the two-point angular correlation for a deep radio sample.
We find that the correlation
function between $0.02^{\circ}$ and $2^{\circ}$ is 
well fitted by a power law of the form 
$A\theta^{\gamma}$ where $A\approx 3\times 10^{-3}$ and $\gamma\approx -1.1$.
On small scales ($\theta<0.2^{\circ}$), double and multi-component sources 
are shown to have a larger clustering amplitude
than that of the whole sample. Sources with flux densities
below 2 mJy are found to have a shallower slope than that obtained 
for the whole sample, consistent with there being a significant 
contribution from starbursting galaxies at these faint fluxes. 
The cross-correlation of
radio sources and Abell clusters is determined. A preliminary 
approach to inferring spatial information is outlined.   
\end{abstract} 

\section{Introduction}
Correlation-function analysis has become the standard way 
of quantifying the clustering of
different populations of astronomical sources. Much of the interest in
this type of analysis stems from its potential to constrain the
spectrum of density fluctuations present in the early universe and thus,
to constrain the physics of the early universe.  
While it is the spatial correlation function that is directly related to
the power spectrum, 
the $angular$ correlation function of galaxies in optical surveys
has been used to infer the spatial correlation function when redshift 
information was not available. These inferred values have later been 
shown to agree with measured values of the
spatial correlation (cf Groth \& Peebles, 1977 and Davis \& Peebles, 1983).  
In addition, in all but the deepest optical
surveys, the amplitude of the angular galaxy-galaxy correlation function
($A$, in $w(\theta) = A\theta^{\gamma}$, $\gamma\approx -0.8$) is found to
scale with the depth of the survey as described in Peebles (1980). Very 
faint galaxies might have a slightly flatter slope, but the fact
that $A$ generally scales in this well-understood way
supports the use of the angular correlation function as a probe of 
large-scale structure (and thus, a probe of the density fluctuation spectrum 
in the early universe.)


Optical surveys covering large regions of sky have provided much 
information on large-scale structure
out to a redshift of $z\approx 0.1$. In addition, `pencil-beam' surveys have 
given some information on structure at redshifts beyond this. Radio surveys,
however,  
generally sample much larger volumes
of space than any optical surveys and so 
have the potential to provide information on much larger 
physical scales. It has been thought that the broad luminosity 
function of
radio sources might, however, wash out any spatial correlations in the 
angular projection. This idea was supported by the work of Webster (1976) 
who determined the angular power spectrum of 
the 4C and GB radio surveys and by the work of Masson (1979) who determined 
the angular correlation 
function for the 6C catalog. The catalogs they used contained only the very
brighest radio sources and neither found any evidence for clustering. More 
recently, however, Peacock \& Nicholson (1991) have shown that bright radio 
sources (flux density, $S>2$ Jy) in a 
narrow redshift band ($0.01<z<0.1$) are spatially
correlated, the sources showing the power-law behaviour 
$\xi=(r/11h^{-1} \rm Mpc)^{-1.8}$. Shaver \& Pierre (1989) also found 
some evidence of clustering towards the supergalactic plane for 
radio sources with $z<0.02$. In addition, marginal
detections of a non-zero angular correlation function have been found for 
sources with $S>35$ mJy in the Green Bank 4.85 GHz survey 
(Kooiman, Burns \& Klypin 1995; Sicotte 1995), suggesting
that spatial correlations will indeed show up in the angular projection if 
a survey goes to faint enough fluxes. 

The FIRST survey is 50 times more sensitive than any previous radio survey
and thus provides an excellent opportunity to investigate the angular
correlation function of faint sources and to explore the possibility of 
inferring spatial information from this measurement. In section 2, we describe
the catalog of $\sim 138,000$ sources used in the study. We then outline 
the methods we have adopted to compute the angular correlation function and the
errors thereon (\S 3). Section 4 presents the results for the 
catalog as a whole, as
well as for various subsamples; in section 5 we compare our results with
previous work and outline a preliminary approach to obtaining information on 
the spatial correlation of radio sources. In section 6 we summarize our 
conclusions.

\section{The Catalog}
The goal of FIRST is to survey, at 1.4 GHz, the 10,000 $\rm deg^{2}$ at the 
north Galactic cap scheduled for 
inclusion in the Sloan Digital Sky Survey. The $5\sigma$ source detection
threshold is $\sim$1 mJy. 
With a positional accuracy better
than $1^{\prime \prime}$ and $\sim 100$ sources per square degree 
confusion is not a problem.  
By October 1995, $\sim 15\%$ of the survey had been completed and a catalog
containing 138,665 sources with $07^{h}16<\alpha< 17^{h}36$ and 
$28^{\circ}.3<\delta<42^{\circ}$ was released (White et al.\ 
1996, hereafter WBGH). Approximately 30\% of these
sources are within $0.02^{\circ}$ of another source and so, for the purposes 
of the correlation function analysis, are considered part of 
double-lobed or multi-component systems. 

The details of the data analysis, map production and catalog generation are 
given in Becker, White and Helfand (1995, hereafter BWH) and in WBGH. In 
brief, images are produced at each point in a grid designed to provide an 
efficient and uniform tiling of the sky. These images are truncated at an 
off-axis angle of $23^{\prime}.5$, weighted and  
summed to yield a set of coadded maps containing $1550\times1150$ 
$1.8^{\prime\prime}$
pixels. The rms noise at each point in the coadded maps is calculated 
from the weighted sum of the measured 
noise values in each of the images that contribute to that point. Sources
with flux densities greater than 5 times the rms noise and a peak flux density 
$\geq$ 1.0 mJy are included in the 
preliminary catalog which is then screened for spurious sources using 
the algorithms 
discussed in BWH. Of the 
138,665 sources, 4,813 were flagged as possible sidelobes. 
Our analysis indicates
that less than 10\% of the flagged sources are real sources, while 
less than 1\% of unflagged sources are spurious (see below).  

BWH estimate the catalog to be 95\% complete to 2 mJy and 80\% complete to
1 mJy. This estimate was inferred from comparison with a deep VLA survey 
containing only 49 sources, but one obtains similar predictions by taking 
the sources detected in the 
Westerbork-Leiden-Berkeley Deep Survey and accounting for the 
the effects of the CLEAN bias, resolution effects, and the adopted peak flux 
density threshold. It appears that only the most extended objects are missing
from the catalog. Completeness studies are discussed further 
in BWH and in WBGH. 

A coverage map has been calculated for the survey, providing effective 
sensitivity estimates with $\sim 4^{\prime}$ resolution. A region is considered
to be covered if the sensitivity at all points in the area is greater than 
0.75 times that at the center of the pointing. This map has been used to
establish the large scale uniformity of the survey. On scales smaller than
the coadded image size ($34.5^{\prime}\times 46.5^{\prime}$), 
BWH claim the variation in 
noise level is less than 15\%, with the sensitivity pattern repeated 
from one field to the next. 
 
\section{Determining the Correlation Function}

The two-point angular autocorrelation function is defined by the 
joint probability
$\delta$P of finding two sources in each of the elements of solid angle 
$\delta \Omega_{1}$ and $\delta \Omega_{2}$ separated by angle $\theta$: 
$$ \delta P = N^{2}[1+w(\theta)]\delta \Omega_{1} \delta \Omega_{2} 
\eqno (1)$$ 
where $N$ is the mean surface density. Many derivations for estimators
of $w(\theta)$ have been given before (see, for example, Peebles 1980).
For those unfamiliar with the field, we outline a basic derivation
given by Sicotte (1995). 

It follows from the definition that for a catalog of $n$ data points 
covering a solid angle $\Omega$, the mean number
of sources at a distance $\theta\pm\Delta\theta$ from a randomly picked 
data point is
$(n-1)(1+w(\theta))\langle\delta\Omega\rangle/\Omega$ where 
$\langle\delta\Omega\rangle$ is the mean solid angle about a randomly
chosen data point. The total number of
pairs with separations in the interval $\theta\pm\Delta\theta$ is then given 
by $DD(\theta)={n\over 2} (n-1)(1+w(\theta))\langle\delta\Omega\rangle/
\Omega$. $DD(\theta)$ can be measured directly from the 
catalog, and when combined with an estimate of $\langle\delta
\Omega\rangle/\Omega$, can give an estimate of $w(\theta)$. For an all-sky
catalog $\delta\Omega=2\pi \sin(\theta)\Delta\theta$ and $\Omega=4\pi$. For
surveys with complicated geometries, it is more practical to calculate 
$\langle\delta\Omega\rangle/\Omega$ using a field of randomly distributed 
points
covering the same area as the survey. Assuming there are the same 
number of random points as there are data points, the number
of pairs of random points with separations between 
$\theta\pm\Delta\theta$ is given by
$RR(\theta)={n\over 2} (n-1)\langle\delta\Omega_{r}\rangle/\Omega$. One can 
clearly measure this quantity and obtain a value for $\langle\delta\Omega_{r}
\rangle/\Omega$
which is approximately equal to $\langle\delta\Omega\rangle/\Omega$.
A simple estimator of $w(\theta)$ is thus given by $DD/RR-1$. Strictly, one
needs to have more random points than data points so that the $w(\theta)$
estimate is not limited by statistical errors in the random points.
An improvement on this estimate of $\langle\delta\Omega\rangle/\Omega$ can
be obtained using the quantity DR($\theta$): the number of 
pairs of points separated by angle $\theta$ where one point is taken from 
the data field and one is from the random field. Using DR instead of RR
enables one to measure the mean solid
angle around data points [$\langle\delta\Omega
\rangle/\Omega$] as opposed to the mean solid angle around random
points [$\langle\delta\Omega_{r}\rangle/\Omega$].

Sicotte (1995) gives a detailed comparison
of the various estimators of $w(\theta)$. We have chosen to quote 
results using the
the standard (S) estimator [$w(\theta)=2DD/DR-1$] as well as the estimator 
suggested by Landy and Szalay (1993, hereafter LS) 
[$w(\theta)=(DD-DR+RR)/RR$]. The LS estimator has been shown to have 
smaller uncertainties on larger scales and Sicotte claims it will remove
the effects of large scale fluctuations in density for estimates of $w(\theta)$
at small $\theta$  . The method introduced by Hamilton (1993) 
[$w(\theta)=4(DD*RR)/(DR*DR)-1)$] gives results very similar 
to the LS estimator. 

To determine the uncertainties associated with each estimate, a 
`bootstrap' analysis of 
the errors was performed. This method of estimating uncertainties is described in
detail in Ling, Frenk and Barrow (1986) and in Fisher et al.\ (1994). A set of 
`bootstrap' catalogs, each the same size as the data catalog, are 
generated using the following procedure. A source is picked at random from the 
data catalog and inserted into the first bootstrap catalog. Random sources
from the data catalog continue to be included in the bootstrap catalog 
until it has the same number of sources as the data catalog.
Some of the sources in the
original data set will appear more than once in the bootstrap catalog, 
some will not appear at all. A set of these bootstrap catalogs can then be
generated and the correlation function can be calculated 
for each one. The result is that at each $\theta$ we produce a set of normally
distributed estimates of the correlation function. The variance around 
the mean can then be used as an estimate 
of the uncertainty in the measurement of $w$ at each $\theta$. This 
process established that both the S estimator and the
LS estimator of $w(\theta)$ had similar 
uncertainties associated with
them out to about $10^{\circ}$. The Poissonian estimate of 
the uncertainties given by $\Delta w = [(w(\theta)+1)]/
[DD(\theta)]^{0.5}$ is less than the bootstrap estimate by a factor of 2 on 
small scales ($\sim 0.05^{\circ}$) and by more than an order of magnitude on 
larger scales ($\sim 5^{\circ}$). 

As explained above, the random field measurements RR and DR 
correct for the `edge effects'; that is, they
estimate the mean solid angle about points in the area of the survey for a 
given $\theta$. Therefore, the random field should contain all the biases 
that the data
field contains. We consider the following:
\begin{itemize}
\item In the correlation analysis, we have used a catalog in which
all sources within $0.02^{\circ}$ of each other are considered a 
single source. This constraint is reproduced in the random fields.

\item The coverage of the data is reproduced in the random fields
using the $2060\times299$ pixel coverage map where each pixel represents a 
$4.5^{\prime}\times3^{\prime}$ ($RA\times dec$) area. Random points are 
generated
in any region where there is data coverage. The flux density threshold
for detection of a source depends on the local rms noise, which varies 
slighlty accross the
survey area. To check whether any systematic variations in sensitivity
could affect the correlation function estimate, we used the coverage map
to generate a random field which also `missed' sources in noisy areas even 
though they had a flux density of $S>1 $mJy. Random sources were assigned
a random flux density according to Windhorst et al.'s (1985) 
log N-log $S$ curve. If the source flux density was below 5 times the 
rms noise given in the coverage map
then it was discarded. The results of this are noted in $\S 4.1$

\item The presence of sidelobe contamination could also be 
included in the random fields. Instead, we decided to use the correlation
function to obtain an independent estimate of the sidelobe contamination.

\end{itemize}	
In addition to the angular autocorrelation function for the catalog as a 
whole and
various source subsamples, we have determined the angular cross 
correlation function for the radio sources and Abell clusters. 
We use the estimator 
$w_{cross}(\theta)= Dd/Rd-1$ where Dd is the number of radio sources 
separated by
$\theta$ degrees from an Abell cluster center and Rd is the number of random 
field points $\theta$ degrees from an Abell cluster center.

\section{Results}
The correlation function fit parameters for various subsamples are given in 
Table 1.
\subsection{The Whole Sample}
A catalog of 109,873 sources was generated by collapsing all sources that are
within $0.02^{\circ}$ of each other to a single source, since the majority
of such cases represent multiple components of a single host object (e.g.,
double radio lobes). This sample predominantly contains 
radio-loud, giant elliptical galaxies, quasars, and
starbursting galaxies (Windhorst et al.\ 1985). 
(Becker et al.\ (1996) have shown that stars make up less than 
0.1\% of the sources.) 
The correlation function of such 
a mixed sample is difficult to interpret but it serves as a starting point
for investigating specific subsamples. The correlation 
function for all sources in the catalog with $07^{h}16<\alpha<17^{h}36$ 
and 
$28.3^{\circ}<\delta<42^{\circ}$ was determined using both the LS estimator
[$w(\theta)=(DD-DR+RR)/RR$] and the S estimator 
[$w(\theta)=2DD/DR-1$]. The RR and DR used are the average of 10 random
field generations. The results for the S estimate are displayed in Figure 1. 
The error bars shown are random errors determined using the bootstrap 
algorithm described above.

To determine the parameters of a power-law fit of the form $A\theta^{\gamma}$,
a straight line was fitted to the log-log plots (for collapsed sources) 
out to $2^{\circ}$ using  
standard $\chi^{2}$ minimization. This yields 
$A=(3.7\pm0.3)\times10^{-3}$ and $\gamma=-1.06\pm0.03$ for the S estimator and 
$A=(2.0\pm0.3)\times10^{-3}$ and $\gamma=-1.26\pm0.04$ for the LS estimator. 
For the LS estimate to be a better
estimate than the S estimate, one requires many more random points in a field
than data points. With about $10^5$ data points, the number of random points
is severely limited by computer time. The S estimate is thus 
probably more reliable.

Also shown in Figure 1 are correlation function estimates obtained when
sources separated by less then $0.02^{\circ}$ are not collapsed. As is
expected, one sees a large increase in the correlation function on small
scales resulting from single sources being counted as two or more sources.
On the scales of our fit, however there is little difference between the 
two, indicating that the small percentage of unassociated sources which
have been merged do not affect the correlation function significantly.
  
There are two problems with fitting a straight line to 
determine power law parameters. The first is that normal errors in the 
original data do not translate to normal errors in the log-log plot as 
required for $\chi^{2}$ fitting. To check
the effects of this, we used the Levenberg-Marquardt technique to fit 
a function of the form $A\theta^{\gamma}$ to the original (linear-linear) 
data. It was found 
that the slopes and amplitudes agreed to well within $1\sigma$ with the 
straight-line fits; we thus consider the straight-line fits to be adequate.
The second problem is that the correlation function estimates at different
$\theta$ are not independent. In calculating the correlation function for
a sample of optical galaxies, Bernstein (1994) used principal component 
analysis to obtain linearly independent combinations of their measurements 
which could then be used in a $\chi^{2}$ fit. It did not appear to affect 
their estimates of the parameters significantly (although the effect
on error estimates can be more substantial).  

As a test of our procedures, random fields were generated and
analyzed as though they were data fields. It was found that the $w(\theta)$ 
were consistent with zero.

Some systematic differences in the uniformity of the survey are evident in the
correlation functions. The result for the sample that 
$includes$ the sources flagged as sidelobes (not plotted) shows a sudden 
increase in 
$w(\theta)$ at $6^{\prime}$. When flagged sources are removed (as in the 
figures shown here), 
the bump decreases in size but is still noticeable, particularly in the 
result for
double and multi-component sources. Six arcminutes is the distance of the 
first sidelobe in uncleaned VLA B-configuration images so the `bump' is a 
clear indicator that not all
sidelobes have been removed. The situation is worse for double and 
multi-component sources because the sidelobes from such extended sources 
are harder to remove in the 
standard cleaning process. 
To investigate further the effect of sidelobes on the correlation function,
varying fractions of spurious sources were added at 
$6^{\prime}$ from real data points. Results indicate that the distortion
in the correlation function is fairly well 
localised to $6^{\prime}$ and that about $\sim1\%$ of all the 
(collapsed) sources remaining in the catalog are sidelobes. Efforts to 
improve the efficiency of our sidelobe flagging algorithms are in progress. 

The first point in the correlation function for the 
whole sample appears high. This could be attributed 
to the presence of double-lobed galaxies that have separations larger
than $0.02^{\circ}$, the adopted radius within which we call all detected 
components a single radio source. 
There is also a significant dip in $w(\theta)$ at $\sim10^{\prime}$. This 
could be related to the cleaning procedure or to the systematic 
non-uniformity in sensitivity that is repeated from one coadded map to 
another. To 
investigate this further we included the sensitivity fluctuations
given in the coverage map in the random field (as described in $\S 3$). This
did result in a smoother estimate at $10^{\prime}$; in addition, the first 
point in the
correlation function was also slightly lower. Fitting a straight line
to the log-log data which included the sensitivity fluctuations returned
parameters which were within $1\sigma$ of those which did not include
sensitivity fluctuations (given above). 

Cross-correlating the whole sample with Abell clusters results in a 
power law fit with an amplitude $\sim 3.6$ times that of the 
autocorrelation (see Figure 2). In the future, we will investigate the 
possibility
of combining this information with spatial information for clusters to 
infer spatial information for radio sources.

\subsection{Double and Multi-component sources}
A catalog of double and multi-component sources (resolved extended sources)  
was generated by considering
any sources separated by less than $0.02^{\circ}$ to be part of a single 
multi-component source. Results for this subsample of 17,773 sources 
using the S estimator are shown for $\theta<0.2^{\circ}$ in Fig.\ 3. 
Force-fitting a line with the same slope as the whole sample 
yields an amplitude of $A=9.5\times 10^{-3}$, a factor of 2.7 times larger 
than that for the sample as a whole. Fitting a straight line on these small 
scales 
yields $A=(3.8\pm0.5)\times10^{-3}$ and $\gamma=-1.4\pm0.4$. On larger
scales many of the error bars extend below zero so we fit the linear-linear
data using the Levenberg-Marquardt technique. This yields 
$A=1.65\times10^{-3}$ and $\gamma=-1.36$ out to $1^{\circ}$, but this does not 
fit the points on small scales very well. 

Traditionally, those radio sources which can be resolved into components have 
been classified into one of two groups (Fanaroff and Riley 1974): 
FR II sources are the more luminous 
($\log P(W \, Hz^{-1})>25$) 
`classical doubles', while FR I sources have distorted lobe structures and 
generally have lower luminosities. FR II sources are known to prefer lower 
density environments than FR I sources at low z, but Hill and Lilly (1991) 
have shown that at z$\sim$0.5 FR II sources also exist in higher density 
regions. In an attempt to estimate the correlation for 
FR II and FR I sources separately, the catalog of multiple sources was split 
into a catalog containing double sources only, and a catalog containing 
sources with more than two components. 
The small number of points resulted in a large amount of scatter, particularly
on larger scales, but on small scales ($\theta<0.2^{\circ}$), it 
appeared that the multi-component sources (predominantly FR I's)
were, as expected, more clustered than the double sources (FR II's) by 
about a factor of 2-3.  

Using the preliminary catalog of sources with $28^{\circ}<\delta<31^{\circ}$ 
that was
available at the beginning of 1995, a catalog of double sources was 
generated using stricter selection criteria than those used here. In addition 
to the
requirement that there be two (and only two) sources separated by less than
$0.02^{\circ}$, the fluxes of the sources were required to be within a 
factor of five of each other. The correlation 
function determined using the more limited sample agrees well with the 
correlation function determined for the sample with the simpler 
selection criteria. 

The cross correlation of Abell clusters with double and multi-component
sources was determined. The correlation amplitude is a factor
of $\sim1.6$ larger than that obtained for the cross-correlation with the
whole sample and $\sim2.5$ larger than that obtained for the 
autocorrelation of this subsample.

\subsection{Flux Cuts and the Presence of Star-Bursting Galaxies}

We have also analyzed a sample that $excluded$ all double and multi-component 
sources. Fitting a straight line, one obtains $A=(5.2\pm0.4)\times10^{-3}$
and $\gamma =-0.84\pm0.05$ using the S estimate and 
$A=(2.2\pm0.4)\times10^{-3}$
and $\gamma=-1.1\pm0.05$ using the LS estimate. These slopes are
shallower than those
obtained for the sample as a whole, and are more consistent with values 
determined from optical surveys. The difference is statistically not very 
significant but it could be related to the fact that this
subsample contains a larger fraction of   
starbursting galaxies as compared to the sample as a whole.
Bright, low-redshift starbusting galaxies are an important component of 
extragalactic
IRAS sources which have been shown to have spatial clustering 
properties similar to those found for optical galaxies (Davis et al.\ 1988). 
The relative contributions of starburst galaxies and AGN to the cumulative
radio source counts is given in BWH as a function of flux density (based on 
Windhorst et al.\ (1985) and Condon (1984)). Below 1.0 mJy
the ratio of the number of starbursting galaxies to the number of AGN 
(including all giant ellipticals) is approaching unity. Above 3 mJy this 
ratio becomes orders of magnitude smaller. To investigate the
contribution of starbursting galaxies to the correlation function, the 
`singles' catalog was divided into catalogs of sources with flux densities 
below 2 mJy and above 3 mJy, respectively. The results are shown in Fig.\ 4.
Best fit parameters
to the $S<$2 mJy sources are $A=(8.2\pm1\times10^{-3})$ and
$\gamma= -0.84\pm0.05$ using the S estimator and $A=(5.6\pm0.6)\times10^{-3}$ 
and 
$\gamma=-0.97\pm0.04$ using the LS estimator. A similar result is obtained for
a sample with 2 mJy$<S<$3 mJy, indicating that incompleteness is not 
a problem here. The shallow slope and large 
amplitude is consistent with there being a larger contribution from `nearby'
starbursting galaxies with clustering properties more similar to optical 
galaxies. 
Best fit parameters
to the $S>$3 mJy sample are $A=(1.9\pm0.4)\times10^{-3}$ and
$\gamma= -1.10\pm0.13$ using the S estimator and $A=(2.6\pm0.8)\times10^{-3}$ 
and $\gamma=-1.2 \pm0.1$ using the LS estimator.
Above 3 mJy the slope is more similar to that obtained 
for all sources, but the amplitude is significantly 
lower. Assuming that clustering does not decrease with 
time, the clustering amplitude of a sample must decrease as the average 
distance to the sources increases. The results are thus consistent with the 
$S>$ 3mJy sources being dominated by more distant, unresolved FR I's and 
FR II's. The lower amplitude could also be related to the presence of a
large fraction of quasars---although radio loud quasars are thought to have
a large clustering amplitude (Bahcall \& Chokski 1991), their 
average distance is larger than that of normal radio
galaxies which will push the clustering amplitude down. 

The whole sample was also divided into various flux density bins.
We determined the correlation function for 3 samples in which all sources 
below 2 mJy, 3 mJy and 10 mJy, were, in turn, excluded. The results were all
similar to that determined for 
the whole survey. This is also true for samples containing sources in 
2-10 mJy and 10-35 mJy flux density bins 
(although the scatter increases as the number
of sources decreases). In contrast, all but the deepest of optical surveys
display a decrease in 
the amplitude of the angular correlation as the limiting magnitude is 
increased (e.g., Maddox et al.\ 1990). A similar result
is not found for radio sources because a lower flux density threshold does not
correspond to a deeper survey as it does for optically selected sources.
The intrinsic luminosities of radio sources vary over many orders of 
magnitude, resulting
in contributions from sources with a wide range of redshifts regardless of
the flux density threshold . In addition, the appearance of starbursting 
galaxies decreases the average redshift as thresholds reach 1 mJy. This effect
is also seen in the deepest optical surveys. The correlation function for 
the 1-2 mJy
flux cut has a slope $\sim 0.85$, consistent with a sample 
containing a significant fraction
of starbursting galaxies with clustering properties more similar to 
optical galaxies.

\section{Discussion}
Calculations of the correlation function for a preliminary catalog 
(October '94) that covered a strip of sky $2.5^{\circ}$ wide in declination 
and $137^{\circ}$ in RA
appear in Cress et al.\ (1995). It was pointed out there 
that the source extraction
algorithms were still being developed and so some changes to the 
calculation would be inevitable, particularly in the case of multiple sources
for which CLEANing is more difficult. An improved version of the
catalog covering this $2.5^{\circ}$ wide strip was later made available at the
FIRST website (January '95). Analysis of this improved
version showed a significant downward shift in the correlation function.
The results for the same strip of sky in the latest catalog (October '95)  
(the catalog used throughout this work and described in Section 2) agree 
well with the results for the January '95 catalog. The
amplitude of the correlation function for the improved catalogs  
is 16\% lower than that for the catalog used in Cress et al.\ (1995).
The result for the double and multi-component sources has changed in shape
and amplitude.

The power-law fits obtained here agree well with 
estimates of the angular correlation function for the Green Bank 4.85 GHz
survey. That catalog contains 54,579 sources with 
$0^{\circ}<\delta<75^{\circ}$ and $S>25$ mJy. 
About 40\% of these were used in correlation function analysis by 
Kooiman et al.\ (1995) and Sicotte (1995).
As a result of the significantly larger number of sources in the FIRST
survey, our random errors are, in some cases, as much as a factor of 10 
smaller than those estimated from Sicotte's 
$\chi^{2}$ contour plots. Our values are within 
his 30\% confidence intervals for the amplitude and slope 
of the correlation of sources with $35$ mJy$<S<900$ mJy. 
The results of Kooiman 
et al.\ are also fairly consistent with ours. 

In 1978, Seldner and Peebles investigated the angular correlation function 
of radio sources in the 4C catalog which contains 4,836 sources with 
$-7^{\circ}<\delta<80^{\circ}$ and $S>2$ Jy. The correlation signal was barely
detectable but they estimated the amplitude of the correlation to be 0.02 at 
$2^{\circ}$. Assuming $\Omega=1$, they derived a relation between 
the angular correlation function of two populations and 
the spatial correlation via the luminosity functions of those populations. 
This enabled them to estimate the amplitude of the spatial correlation 
function of the radio sources (assuming power law behaviour). Their best 
fit gave a spatial correlation scale of $r_{0}h^{-1}\sim74$ Mpc. 
Writing their equation (27) for the
autocorrelation of radio sources and substituting for $N$, the number of 
sources per steradian brighter than flux density $S$, one obtains:
$$w(\theta)={{\int_{0}^{\infty}x^{4}dx \phi^{2}
\int_{-\infty}^{+\infty}dy \xi[r,z(x)]}\over{[\int_{0}^{\infty} x^{2} dx \phi[\log P,z(x)]]^2}} \eqno (2)$$
where the coordinate distance is given by 
$$x={2c\over{H_{0}a_{0}}}[1-(1+z)^{-0.5}]\eqno (3)$$
and the difference in coordinate distance between two sources separated by 
distance $r$ is given by $y$. 
It is assumed that $y<<x$; that is, that clustering is only appreciable on 
scales much less than the distance of the sources.
The selection function is given by 
$$\phi=\int_{\log (P_{min}({\it S}))}^{\log (P_{max})}\rho(\log P,z)d(\log P)\eqno (4)$$
where $\rho$ is the density of radio sources at redshift z with 
luminosity between $\log P$ and $\log P+d(\log P)$ at 1.4 GHz.
$P_{max}$ is given by the maximum luminosity expected 
($\rm \log P(W\, Hz^{-1}sr^{-1})\sim30$) 
for radio
sources. $P_{min}$ is the minimum luminosity required for a source to be seen
at redshift z. This is given by $$P_{min}=Sa_{0}^{2}x^{2}(1+z)^{1+\alpha} 
\eqno (5)$$
where $\alpha$ is the spectral index of the source.  


Dunlop \& Peacock (1991) (DP) have estimated the radio luminosity function 
(RLF) at 2.7 GHz for steep spectrum and flat spectrum sources separately. 
Condon (1984) has estimated the RLF at 
1.4 GHz. Their estimates are derived from source counts at various 
wavelengths, from redshifts of only the brighest sources (S$>\sim$1 Jy)
and from photometry of some fainter sources. In a first attempt at comparing 
Peacock and Nicholson's spatial 
correlation (see $\S 1$) with that inferred from our measurements of 
the angular 
correlation, we assume $\xi(r,z)=(r/r_o)^{\gamma-1}(1+z)^{-3.8}$. The redshift
dependence corresponds to linear growth of density perturbations. Following 
Seldner and Peebles, we can then estimate the value of $r_0$ using 
Condon's RLF or DP's combined steep and flat RLF's, translated to 1.4 GHz.

We calculated $r_0$ for $\gamma=-0.8$ and $\gamma=-1.1$ using both DP's 
model and Condon's model. We estimate $r_0$ to be $10h^{-1}$Mpc
but this result is rather sensitive to the assumed clustering
evolution. A detailed analysis of the results will be 
presented in a later paper (Cress \& Kamionkowski 1997).

\section{Conclusions}
\begin {itemize}
\item We have shown conclusively that spatial clustering of radio sources 
can be
detected in the angular projection using the FIRST survey catalog. 
Specifically, we find that the angular correlation function
for all sources above 1 mJy shows power law behaviour with 
$\gamma \sim -1.1$ and $A\sim 3\times 10^{-3}$ out to $\sim4^{\circ}$.
\item Double and multi-component sources (FR II's and FR I's) show,
on scales less than $0.2^{\circ}$, a
correlation amplitude which is larger than that obtained for the whole sample
 by a factor of 2 to 3. Fits on these small scales, however, do not seem to 
extrapolate successfully to larger scales. 
\item The correlation function of the faintest sources ($<$2 mJy) 
appears to have a slope that is significantly shallower than  
that for the brighter sources, suggesting a significant
contribution from starbursting galaxies at these faint fluxes.
 
\end{itemize}   

In the future, we will be refining the subsample selection using both source 
morphology and data on optical counterparts and we will be continuing to
investigate the derivation of spatial information using luminosity functions
and cross-correlations with other populations. 
This should allow us to probe
source clustering to much higher redshifts than has previously been possible.
    
\acknowledgements
The success of the FIRST survey is in large measure due to the assistance
of a number of organizations. In particular, we acknowledge support from
NRAO, NSF (grants AST-94-19906 and AST-94-21178), IGPP/LLNL, the STScI, the
National Geographic Society(NGS No. 5393-094), Columbia University, and
Sun Microsystems. This is contribution Number 607 of the Columbia Astrophysics
Laboratory.

\clearpage
\noindent
FIGURE CAPTIONS

\bigskip

FIGURE 1: The squares show the autocorrelation function ($w(\theta)=2DD/DR-1$)
calculated for the 
whole sample (109,873 sources, where sources separated by less than 
$0.02^{\circ}$ have been collapsed to a single source). The triangles show
the correlation function obtained when sources are not collapsed. 
Error bars are 
obtained using the bootstrapping technique described in Section 3. 
\medskip

FIGURE 2: The cross-correlation function of the whole sample of radio sources
with Abell clusters using the $Dd/Rd-1$ estimator. The solid line shows the
best fit to the cross-correlation results while the dotted line shows the
best fit to the auto-correlation of the radio sources (S) estimator. Error
bars show Poissonian errors ($\Delta w=(w+1)/ \surd DD$)
\medskip
 
FIGURE 3: Open squares show the autocorrelation function for the double 
and multi-component systems using the S estimator with the 
solid line showing the best fit to the points. The filled squares and dashed 
line are the same as those shown in Figure 1. 
\medskip

FIGURE 4: The autocorrelation function of two subsamples: the circles show
the result for all single-component sources with flux densities $<$ 2 mJy, the
stars show the result for single component sources with flux densities 
$>$3 mJy.

\bigskip
\bigskip

\begin{table*}
\caption{Correlation Analysis Results}
\begin{center}
\begin{tabular}{cccccc}
Sample & Source Type & Source Number & Amplitude ($\times10^{-3}$) & Power($\gamma$) \\
\tableline

all sources & mixed & 109,873 & $3.7\pm0.3$ & $-1.06\pm0.03$ \\
double+multi & FR I \& II & 17,773 & $3.8\pm0.5$* & $-1.39\pm 0.4$\\
single & UFR, SB \& C & 92,057 & $5.2\pm0.4$ & $-0.84\pm0.05$\\
single, S$>$3mJy & UFR, C \& SB & 33,800 & $1.9\pm0.4$ & $-1.10\pm0.07$\\
single, S$<$2mJy & SB, UFR \& C &  45,312 & $8.2\pm1.0$ &
$-0.84\pm0.05$\\
all$\times$Abell & mixed$\times$clusters & 109,873$\times$382 & $13.5\pm4.8$ & $-1.07\pm0.2$\\ 
\end{tabular}
\end{center}
\tablenotetext{}{Fitted parameters for various subsamples of data. 
Values are obtained 
from the $(DD/DR-1)$ method of estimation. Fitting is done out to $2^{\circ}$
except for the one labled with *, which is fitted to $0.2^{\circ}$. C refers 
to `compact' sources, SB refers 
to starburst galaxies and FR refers to the Fanaroff-Riley classification. U
stands for unresolved.}
\end{table*}

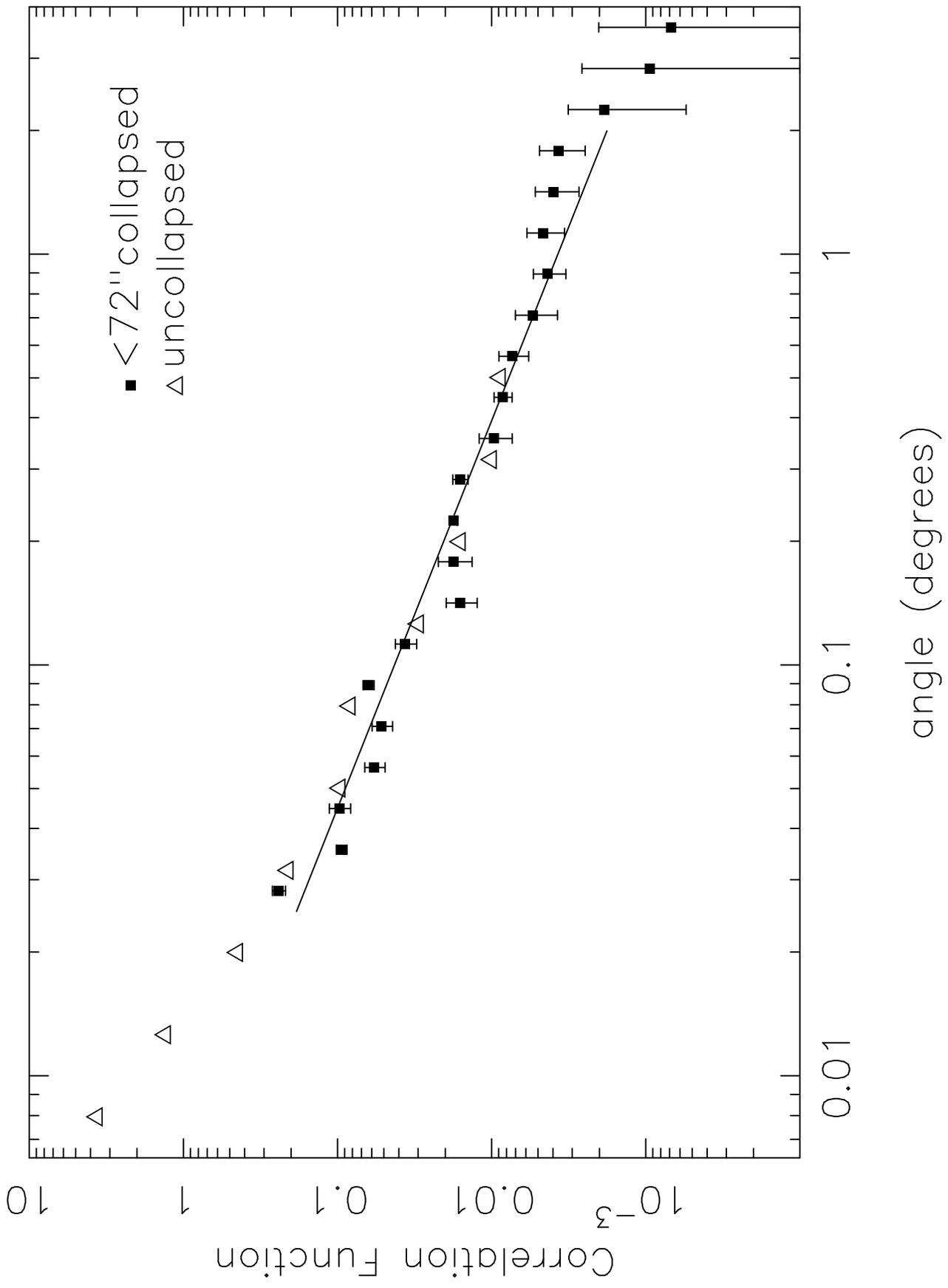
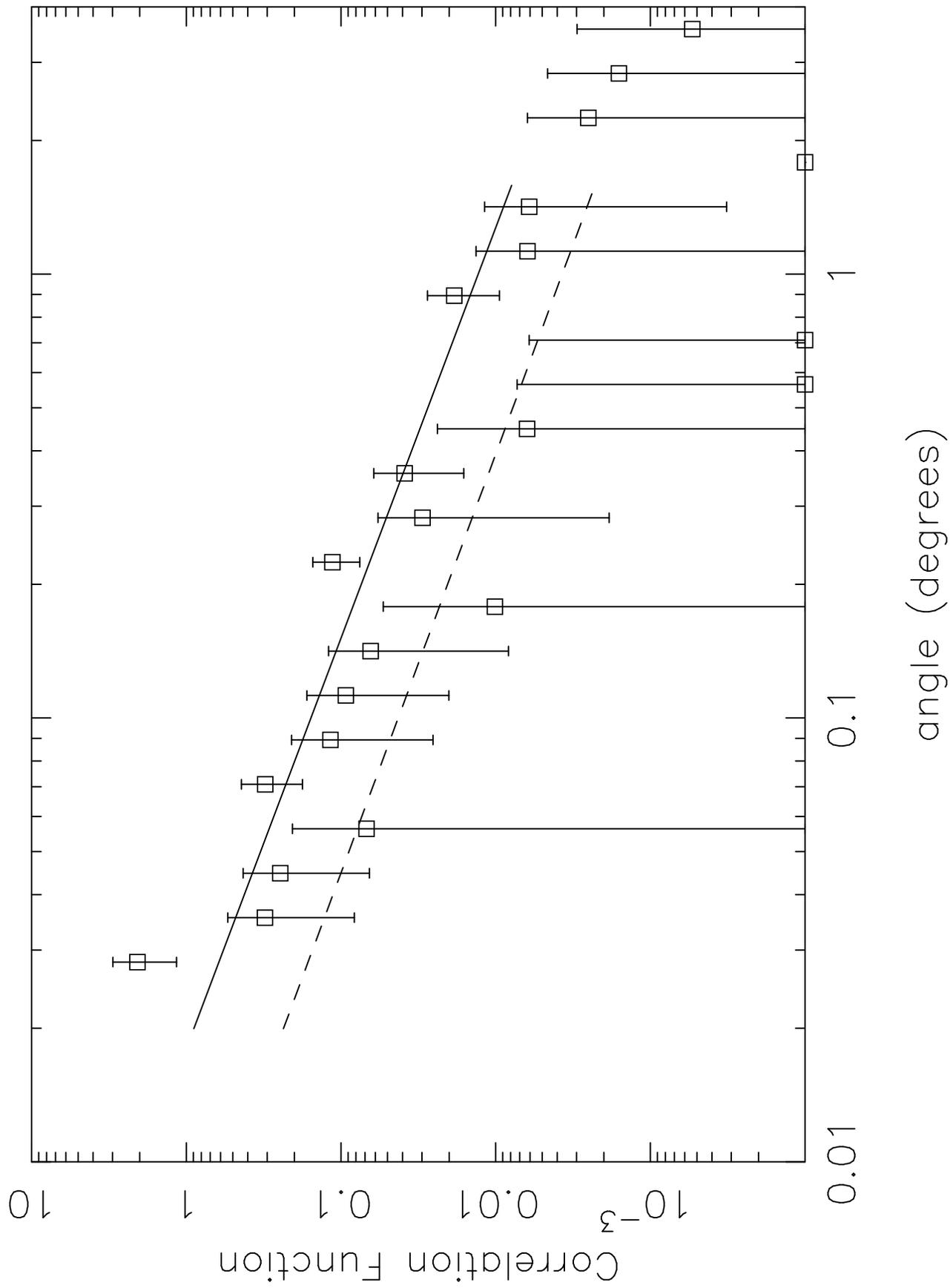
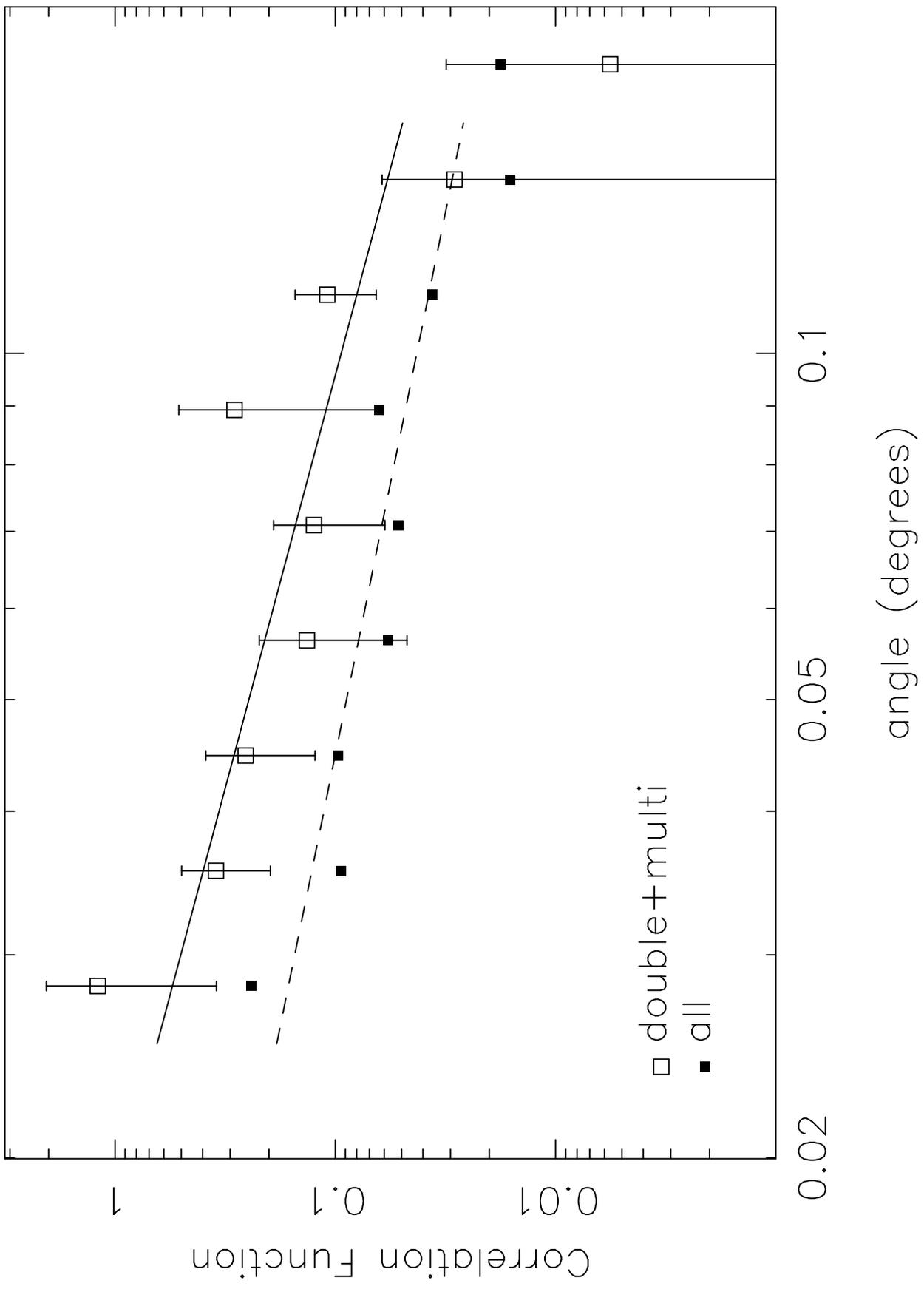
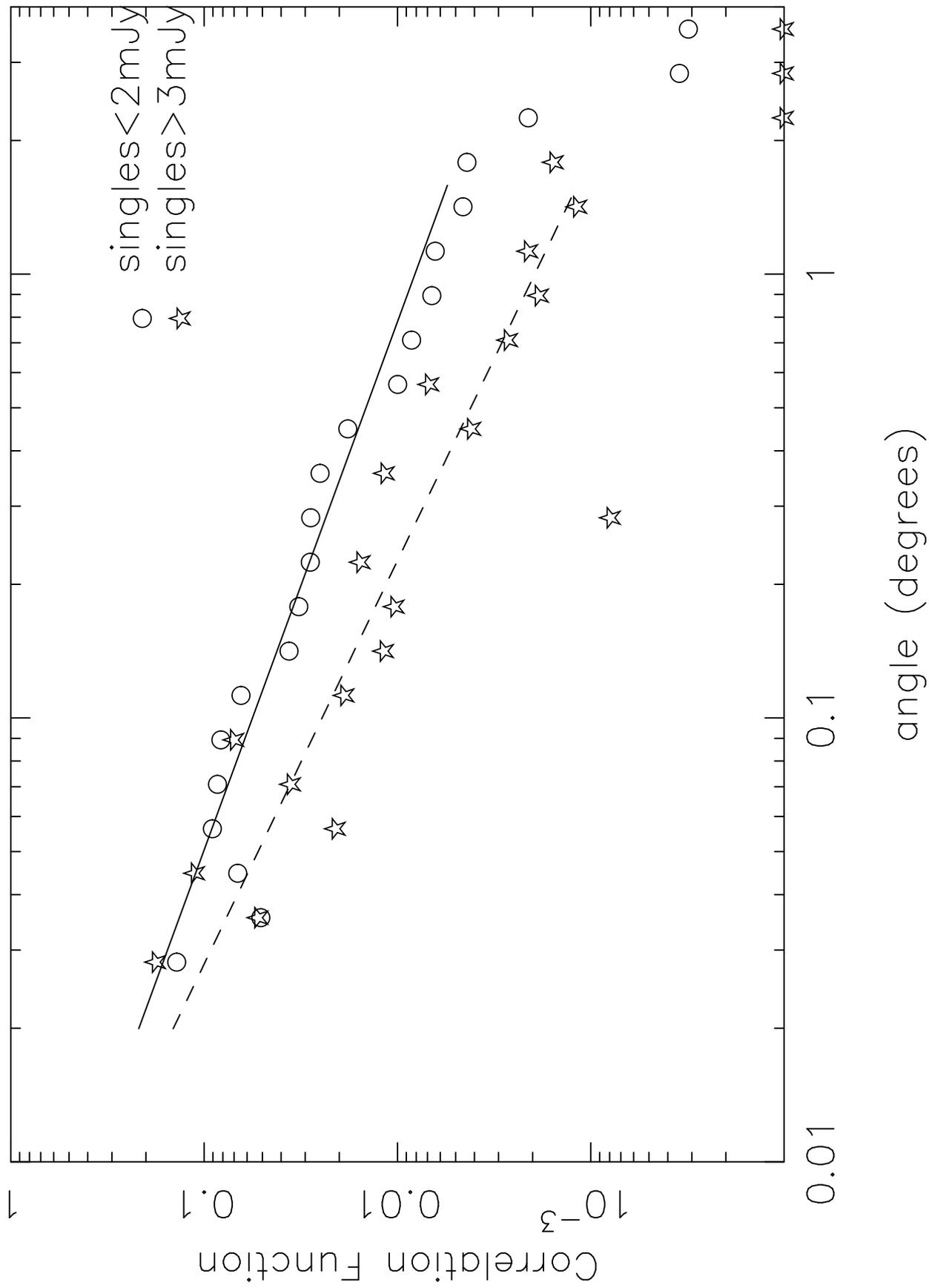
\end{document}